\documentstyle[aps,prb,epsfig,float,twocolumn]{revtex}
\begin{document}
\twocolumn[\hsize\textwidth\columnwidth\hsize\csname
@twocolumnfalse\endcsname

\title{Disorder-induced noncollinear ferromagnetism in models
for (III,Mn)V semiconductors}

\author{John Schliemann}

\address{Department of Physics and Astronomy, University of Basel, 
CH-4056 Basel, Switzerland}

\address{Department of Physics, The University of Texas, 
Austin, TX 78712}

\date{\today}

\maketitle

\begin{abstract}
We study the ground state properties of kinetic-exchange models for
(III,Mn)V semiconductors with randomly distributed Mn ions. Our method
is embedded in a path integral spin-wave type
formalism leading to an effective action for Mn spins only with full 
Matsubara frequency dependence. The zero-frequency contribution to
this action is equivalent to static perturbation theory and characterizes
the stability of a given spin configuration, while the component linear in 
frequency can be interpreted as the joint Berry phase of the Mn and carrier
system.

For simple parabolic-band carriers the collinear ferromagnetic state 
with all Mn spins in parallel is always {\em stationary} but generically
{\em unstable}. This instability can be characterized in terms of inverse
participation ratios and is due to long-ranged nonlocal spin fluctuations.
We also present results for the ground state magnetization as a function
of an external field.

For carrier dispersions involving anisotropy induced by spin-orbit coupling
the collinear state is not even stationary and therefore also not the
ground state. This interplay between the anisotropy in the
carrier system and the disorder in the Mn positions reflects recent findings
by Zarand and Janko (Phys. Rev. Lett. {\bf 89}, 047201 (2002))
obtained within the RKKY approximation.
The stationarity of the collinear state (with the magnetization pointing in one
of the cubic symmetry directions) is restored in the continuum or
{\em virtual crystal} approximation where disorder is neglected. 
\end{abstract}
\vskip2pc]


\section{Introduction}

Since several years diluted ferromagnetic semiconductors are in the focus of 
research in solid state physics; for reviews see 
\cite{Furdyna88,Ohno98,Ohno99,Ohno01,Konig01a,Dietl02,Lee02,Sanvito02}.
This is on the one hand because these systems provide interesting and
challenging physical problems on their own right. On the other hand this
big deal of interest is due to possible applications of these materials in
the emerging field of spintronics \cite{Wolf01,Awschalom02} 
since they offer the perspective of combining ferromagnetism with the 
readily tunable transport properties of semiconductors.

An important achievement in this field was the fabrication of diluted 
Mn-doped GaAs via low temperature molecular beam epitaxy by Ohno and 
collaborators in 1996 \cite{Ohno96}. This material showed a Curie temperature
of 110K, a result which has been reproduced in the meantime by several
other groups. Moreover, very recently reports on Mn-doped (III,V) 
semiconductors having Curie temperatures of room temperature or higher
have appeared, and also other combinations of magnetic ions and/or host
materials look prospective \cite{Reed01,Sonoda01,Theodoropoulou01,Theodoropoulou02,Park02,Ueda01,Matsumoto01,Okabayashi02}.

These recent developments have also already generated  a large amount of 
theoretical research on diluted ferromagnetic semiconductors
\cite{Konig01a,Dietl02,Lee02,Sanvito02,Dietl97,Dietl00,Dietl01a,Jungwirth99,Lee00,Konig00,Abolfath01,Konig01b,Schliemann01a,Schliemann01b,Schliemann02,Jungwirth02,Yang02,dosSantos02,Fernandez-Rossier01,Vurgaftman01,Bouzerar02,Berciu01,Kennett02a,Kennett02b,Zarand01,Chudnovskiy02,Kaminski02,Timm02,Calderon02,Mayr02,Bhattacharjee00,Akai98,Korzhavyi02,Litvinov01,Dietl01b,Schilfgaarde01,Sanvito01,Sato02}. 
Such studies include electronic structure calculations based on 
density functional techniques \cite{Sanvito02,Schilfgaarde01,Sanvito01,Sato02},
and investigations using specific models for such systems
\cite{Konig01a,Dietl02,Lee02,Dietl97,Dietl00,Dietl01a,Jungwirth99,Lee00,Konig00,Abolfath01,Konig01b,Schliemann01a,Schliemann01b,Schliemann02,Jungwirth02,Yang02,dosSantos02,Fernandez-Rossier01,Vurgaftman01,Bouzerar02,Berciu01,Kennett02a,Kennett02b,Zarand01,Chudnovskiy02,Kaminski02,Timm02,Calderon02,Mayr02,Bhattacharjee00,Akai98,Korzhavyi02,Litvinov01,Dietl01b}.
In particular, very recently a series of studies has appeared on 
disorder-related phenomena in different models for diluted ferromagnetic 
semiconductors 
\cite{Schliemann01b,Schliemann02,Jungwirth02,Yang02,Bouzerar02,Berciu01,Kennett02a,Kennett02b,Zarand01,Chudnovskiy02,Kaminski02,Timm02,Calderon02,Mayr02,Bhattacharjee00,Akai98,Korzhavyi02}.
In the present work we employ kinetic-exchange models 
for carrier-mediated ferromagnetism occurring in Mn-doped (III,V)
semiconductors which have been the basis of a large body of previous work
\cite{Konig01a,Dietl97,Dietl00,Dietl01a,Jungwirth99,Lee00,Konig00,Abolfath01,Konig01b,Schliemann01a,Schliemann01b,Schliemann02,Jungwirth02,Yang02,dosSantos02,Fernandez-Rossier01}. The general
Hamiltonian reads
\begin{equation}
{\cal H}={\cal H}_{kin}
+\sum_{I}\int\,d^{3}r\,
J(\vec r-\vec R_{I})\vec s(\vec r\, ) \cdot \vec S_{I}\,.
\label{defmod1}
\end{equation}
These models show several features in accordance with experiments.
They  consist of a kinetic term 
${\cal H}_{kin}$ for free valence-band carriers (holes)
whose spin density $\vec s(\vec r\, )$
is antiferromagnetically exchange-coupled to localized
Mn spins $\vec S_{I}$ of length $S=5/2$
at locations $\vec R_{I}$ by an spatially extended 
coupling $J(\vec r\, )$, which we take to be of the form
\begin{equation}
J(\vec r \,)=\frac{J_{pd}}{(2\pi a_{0}^{2})^{\frac{3}{2}}}
e^{-\frac{r^{2}}{2 a_{0}^{2}}}\,.
\label{excpl}
\end{equation} 
The regularization parameter $a_{0}$ reflects the spatial 
range of the exchange coupling \cite{Bhattacharjee00}.
In a minimal description of carrier-induced ferromagnetism the kinetic term
represents just a simple parabolic band characterized by an effective mass
$m^{\ast}$. 
Moreover we will also consider the case
of a more realistic $\vec k\cdot\vec p$ Hamiltonian \cite{Luttinger55,Chow99,Abolfath01} describing the valence band structure of (III,V) semiconductors
such as GaAs.

Differently from earlier work \cite{Dietl97,Dietl00,Dietl01a,Jungwirth99,Lee00,Konig00,Abolfath01,Konig01b,Jungwirth02}we will not make use
of the {\em virtual crystal} approximation, i.e. we will not approximate
the localized Mn moments by a continuum but retain them as individual
and randomly distributed spins. This additional feature of the model
grossly enriches its physical properties and gives rise to the
occurrence of {\em noncollinear ferromagnetism} \cite{Schliemann02}.
This noncollinearity in the orientations of localized magnetic moments even 
in the ground state of the system is very likely to be an ingredient
to the interpretation of recent experiments by Potashnik {\em et al.}
\cite{Potashnik01}
who found a strong dependence of the magnetic properties of Mn-doped GaAs
on the annealing history of the sample. Similarly strong dependencies are also
found in transport measurements \cite{Potashnik01} and in crystallographic
properties \cite{Schott01}.

Our theoretical method and results to be presented here can be embedded in
a path integral approach to the partition function of the underlying model.
A part of the results was already discussed in 
\cite{Schliemann02,Konig01a}. In the present paper we add further details 
of the formalism and present new results on the frequency dependence of the
action kernel and the role of an external magnetic field.

Moreover, we give a detailed discussion of the gradient of the carrier 
ground state energy with respect to fluctuations in the orientations of the
localized Mn moments. For the case of isotropic systems with simple
parabolic-band carriers the collinear ferromagnetic state with all Mn spins
in parallel is always {\em stationary} (i.e. it has a vanishing energy 
gradient),
but in general not {\em stable} \cite{Schliemann02}. In the present work
we show the possibility of {\em complanar}
spin structures which are local minima on the energy landscape of such systems.
For the case of models with spin-orbit anisotropy in the carrier system
we find that the collinear state is not even stationary and does therefore
not represent an energy minimum. This finding resembles recent results
by Zarand and Janko \cite{Zarand01}. 

Another study related to the
present one was reported on recently by Korzhavyi {\em et al.}
\cite{Korzhavyi02}. These authors investigate, by both experimental
and theoretical means, the influence of As antisite
defects on the ground state structure of the Mn local magnetic moments.
As a result, As antisite defects are found to favor the formation of 
disordered local moment configurations of Mn spins with antiparallel
orientation and therefore reduce the net magnetization. However, differently
from the results of the present investigations, the Mn spins
in the disordered local moment configurations are restricted to 
have (at random) either spin up or down
with respect to some given quantization axis, i.e. they are collinear
to each other.

This paper is organized as follows. In section \ref{gentheo} we
review our general theory of magnetic fluctuations in the
kinetic exchange models studied here, and add further technical details.
In section \ref{inst} we report on numerical results concerning the instability
of the collinear ferromagnetic state in the parabolic-band model.
In section \ref{noncollmag} we discuss several results arising from the
very general gradient expression for the ground state energy obtained in 
section \ref{gentheo}. In subsection \ref{helical} we demonstrate the 
occurrence
of {\em complanar} or {\em helical} energy minima in the parabolic-band
model. An important further finding is discussed in 
\ref{nonstatsixband} where the collinear state is shown 
{\em not even to be stationary} with respect to magnetic fluctuations 
if anisotropy induced by spin-orbit coupling
is present. In subsection \ref{stepdesc} we report on
numerical steepest-descent studies of the true energy minima
below the collinear state, and on the influence of an external magnetic
field. In section \ref{finfreq} we discuss the small-frequency dependence
of the effective-action kernel derived in \ref{gentheo}.
We close with conclusions in section \ref{conclusions}.


\section{Magnetic fluctuations: General theory}
\label{gentheo}

\subsection{Path integral approach}

We now describe a path integral approach to the partition function
of fluctuations around a given magnetic state of the Mn spins. 
Our method is similar to the one used in Refs. \cite{Konig00}
with the differences that we do not approximate the Mn magnetic moments
by a continuum but retain them as individual localized spins, and we
study fluctuations around more general noncollinear magnetic states.
The state we consider is a tensor product of spin-coherent states
for each Mn location $I$ with a polarization direction
$\vec \Omega_{I}=(\sin\vartheta_{I}\cos\varphi_{I},
\sin\vartheta_{I}\sin\varphi_{I},\cos\vartheta_{I})$. We introduce local
coordinates for each Mn site whose $\hat z$-direction coincides with
$\vec \Omega_{I}$. The spin operators $\vec T_{I}$ with respect to these
local bases are related to the original operators $\vec S_{I}$ via
\begin{eqnarray}
\left(
\begin{array}{c}
S^{x}_{I}\\
S^{y}_{I}\\
S^{z}_{I}
\end{array}
\right)
& = &
\left(
\begin{array}{ccc}
\cos\varphi_{I} & -\sin\varphi_{I} & 0 \\
\sin\varphi_{I} & \cos\varphi_{I} & 0  \\
0 & 0 & 1 \\
\end{array}
\right)\nonumber\\
& \cdot &   
\left(
\begin{array}{ccc}
\cos\vartheta_{I} & 0 & \sin\vartheta_{I} \\
0 & 1 & 0 \\
-\sin\vartheta_{I} & 0 & \cos\vartheta_{I} \\
\end{array}
\right)
\left(
\begin{array}{c}
T^{x}_{I}\\
T^{y}_{I}\\
T^{z}_{I}
\end{array}
\right)\,.
\end{eqnarray}
To parametrize fluctuations around the magnetic state given by the directions
$\vec \Omega_{I}$ we use the usual Holstein-Primakoff representation 
of the spin operators $\vec T_{I}$,
\begin{equation}
T^{+}=\sqrt{2S-b^{+}_{I}b_{I}}b_{I}\quad,\quad T^{z}=S-b^{+}_{I}b_{I}\,.
\end{equation}
Formulating the partition function as a path integral and representing the
Mn spin Holstein-Primakoff bosons in a coherent-state
parameterization one arrives after integrating out the carrier degrees of
freedom at the following effective action:
\begin{equation}
S_{\rm eff}=\int_{0}^{\beta}d\tau\sum_{I}
\left[\bar z_{I}\partial_{\tau}z_{I}\right]
-{\rm tr}\left[\ln\left(G^{-1}_{MF}+\delta G^{-1}\right)\right]\,.
\label{fullaction}
\end{equation}
$z_{I}(\tau)$ stands for the bosonic Holstein-Primakoff field 
parameterizing the fluctuations of the Mn spin $I$ around its
mean direction $\vec\Omega_{I}$.
The integration over the imaginary time $\tau$ goes from zero to the
inverse temperature $\beta$, and the trace in the second contribution is over
fermionic carrier degrees of freedom and imaginary time. 
The the fluctuation-free (i.e. mean-field) part 
$G^{-1}_{MF}$ of the integral kernel is given by
\begin{eqnarray}
G^{-1}_{MF}  & = & \partial_{\tau}-\mu-{\cal H}_{kin}
\nonumber\\
& + & S\sum_{I}J(\vec r-\vec R_{I})
\Big[\cos\vartheta_{I}s^{z}\nonumber\\
& & +\frac{1}{2}\left(e^{-i\varphi_{I}}\sin\vartheta_{I}s^{+}
+e^{+i\varphi_{I}}\sin\vartheta_{I}s^{-}\right)\Big]
\end{eqnarray}
where $\mu$ is a chemical potential and  $s^{z}$ and $s^{\pm}=
s^{x}\pm is^{y}$ are carrier spin operators. In the case of a simple parabolic 
band they are just proportional to Pauli matrices while, for instance,
in the case of a six-band $\vec k\cdot\vec p$ Hamiltonian
they have a more complex form \cite{Luttinger55,Abolfath01}.

The fluctuation part of the inverse Green's function in up to quadratic order
in the Holstein-Primakoff variables 
reads $\delta G^{-1}=\delta G^{-1}_{1}+\delta G^{-1}_{2}$ with
\begin{eqnarray}
\delta G^{-1}_{1} & = & \frac{1}{2}
\sum_{I}J(\vec r-\vec R_{I})\nonumber\\
& \cdot\Biggl[ & e^{-i\varphi_{I}}\sqrt{2S}
\left(\bar z_{I}\cos^{2}\frac{\vartheta_{I}}{2}
-z_{I}\sin^{2}\frac{\vartheta_{I}}{2}\right)s^{+}\nonumber\\
& + & e^{+i\varphi_{I}}\sqrt{2S}
\left(z_{I}\cos^{2}\frac{\vartheta_{I}}{2}
-\bar z_{I}\sin^{2}\frac{\vartheta_{I}}{2}\right)s^{-}\nonumber\\
& - & \sqrt{2S}\sin\vartheta_{I}\left(z_{I}+\bar z_{I}\right)
s^{z}\Biggr]\,,
\label{delta1}
\end{eqnarray}
\begin{eqnarray}
\delta G^{-1}_{2} & = & -
\sum_{I}J(\vec r-\vec R_{I})
\Big[z_{I}\bar z_{I}\cos\vartheta_{I}s^{z}\nonumber\\
& +& \frac{1}{2}z_{I}\bar z_{I}\sin\vartheta_{I}
\left(e^{-i\varphi_{I}}s^{+}+ e^{+i\varphi_{I}}s^{-}\right)\Big]\,.
\label{delta2}
\end{eqnarray}

To analyze magnetic fluctuations we expand the action
(\ref{fullaction}) in the bosonic spin variables describing deviations
from the prescribed directions $\vec\Omega_{I}$,
\begin{eqnarray}
{\rm tr}\ln\left(G^{-1}_{MF}+\delta G^{-1}\right) & = & 
{\rm tr}\ln\left(G_{MF}\right)\nonumber\\
&  & -\sum_{n=1}^{\infty}
\frac{1}{n}{\rm tr}\left(-G_{MF}\delta G^{-1}\right)^{n}\,.
\label{actionexpansion}
\end{eqnarray}
We will evaluate the effective action in up to second oder in the fluctuations
$z_{I}(\tau)$. To this end we introduce the Fourier transforms
$z_{I}(\Omega_{n})=\int_{0}^{\beta}d\tau \exp(i\Omega_{n}\tau)z_{I}(\tau)$ 
with the notation $\bar z_{I}(\Omega_{n})={\overline{z_{I}(\Omega_{n}))}}$, 
where $\Omega_{n}=2n\pi/\beta$, $n$ 
integer, is a bosonic Matsubara frequency.

In this subsection we shall consider a simple parabolic-band model where
fluctuation-free part of the carrier Green's function reads in real-space 
representation
\begin{equation}
G_{MF}(\vec r,\sigma;\vec r',\sigma';\omega_{n})
=-\sum_{\alpha}
\frac{\psi_{\alpha\sigma}(\vec r){\bar\psi}_{\alpha\sigma'}(\vec r')}
{i\omega_{n}-\eta_{\alpha}}\,.
\end{equation}
Here $\omega_{n}=(2n+1)\pi/\beta$, $n$ integer,
is a fermionic Matsubara frequency and
$\psi_{\alpha\sigma}(\vec r)$ is the spin component $\sigma$ of the 
carrier wave function with label $\alpha$ and energy $\varepsilon_{\alpha}=
\eta_{\alpha}+\mu$. These wave functions are the eigenstates of the
single-particle mean-field Hamiltonian
${\cal H}_{MF}=G^{-1}_{MF}-\partial_{\tau}+\mu$.

The lowest-order contribution in the effective action
occurs at zero Matsubara frequency only and 
is {\em linear} in the Holstein-Primakoff variables:
\begin{equation}
{\cal S}_{\rm fluc}^{(1)}=\frac{1}{2}\sum_{I}
\left[\bar g_{I}z_{I}(0)+g_{I}\bar z_{I}(0)\right]
\label{gradientfluc}
\end{equation}
with $g_{I}=g^{1}_{I}+ig^{2}_{I}$, and
\begin{eqnarray}
g^{1}_{I} & =  & \sqrt{2S}\left(\vec e_{\varphi_{I}}\times\vec e_{z}\right)
\cdot\int d^{3}r\Bigl[J(\vec r-\vec R_{I})\nonumber\\
 & & \Bigl(\left(\langle\vec s(\vec r)\rangle\cdot\vec e_{\varphi_{I}}\right)
\vec e_{\varphi_{I}}+
\left(\langle\vec s(\vec r)\rangle\cdot\vec e_{z}\right)\vec e_{z}
\Bigr)\times\vec \Omega_{I}\Bigr]
\label{g1}\,,\\
g^{2}_{I} & =  &\sqrt{2S}\vec e_{z}\cdot
\left(\vec e_{\varphi_{I}}\times\int d^{3}rJ(\vec r-\vec R_{I})
\langle\vec s(\vec r)\rangle\right)\,.
\label{g2}
\end{eqnarray} 
Here $\langle\vec s(\vec r)\rangle$ is the expectation value of the
carrier spin density,
$\vec e_{\varphi_{I}}=(\cos\varphi_{I},\sin\varphi_{I},0)$ and
$\vec e_{z}=(0,0,1)$.

The contribution $S_{\rm fluc}^{(2)}$ bilinear in the Holstein-Primakoff
variables is quite complex for general Mn spin orientations $\Omega_{I}$.
In particular, numerical evaluations of this quantity are extraordinarily
tedious and computationally expensive. We therefore shall concentrate
on the collinear state where all Mn spins point, say, along the $z$-direction.
Then the second-order contribution to the effective action reads
\begin{equation}
S_{\rm fluc}^{(2)}=\frac{1}{\beta}\sum_{n}\sum_{I,J}
\bar z_{I}(\Omega_{n})D^{-1}_{IJ}(\Omega_{n})z_{J}(\Omega_{n})
\label{fluctuation}
\end{equation}
where the fluctuation matrix $D^{-1}_{IJ}(\Omega_{n})$ reads
\begin{equation}
D^{-1}_{IJ}(\Omega_{n})=L_{IJ}(\Omega_{n})+K_{IJ}(\Omega_{n})
\end{equation}
with
\begin{eqnarray}
L_{IJ} & = &\delta_{IJ}\Big(-i\Omega_{n}-
 \int d^{3}rJ(\vec r-\vec R_{I})\langle s^{z}(\vec r)\rangle\Big)
\label{L}\,,\\
K_{IJ} & = & \frac{S}{2}\sum_{\alpha,\beta}\Biggl[
\frac{n_{F}(\eta_{\alpha})-n_{F}(\eta_{\beta})}
{i\Omega_{n}+\eta_{\alpha}-\eta_{\beta}}
F^{\alpha\downarrow,\beta\uparrow}_{I}
F^{\beta\uparrow,\alpha\downarrow}_{J}\Biggr]\,.
\label{K}
\end{eqnarray}
Here $n_{F}$ is the Fermi function, and 
\begin{equation}
F^{\alpha\sigma,\beta\mu}_{I}=\int d^{3}rJ(\vec r-\vec R_{I})
\bar\psi_{\alpha\sigma}(\vec r)\psi_{\beta\mu}(\vec r)\,.
\label{F}
\end{equation}
All quantities referring to the carrier system are to be
evaluated for the collinear orientation of Mn spins.

The diagonal contributions to the action kernel summarized in 
$L_{IJ}$ stem from the kinetic term 
$\sum_{I}\bar z_{I}\partial_{\tau}z_{I}$ in the integrand of 
(\ref{fullaction}) and from the lowest-order term
${\rm tr}(G_{MF}\delta G^{-1}_{2})$ in (\ref{actionexpansion}).
The term given by $K_{IJ}$ arise from the bubble
contribution $-({\rm tr}(G_{MF}\delta G^{-1}_{1}G_{MF}\delta G^{-1}_{1}))/2$.

\subsection{The static limit: Perturbation theory}
\label{pert}

The limit of zero Matsubara frequency $\Omega_{n}$
corresponds to static perturbations of the  mean-field carrier ground state.
It is instructive to verify this explicitly by elementary perturbation 
theory. The unperturbed mean field Hamiltonian is again
${\cal H}_{MF}=G^{-1}_{MF}-\partial_{\tau}+\mu$ which is subject to a 
perturbation ${\cal H}_{fluc}[\{z_{I},\bar z_{I}\}]$ with
\begin{equation}
{\cal H}_{fluc}[\{z_{I},\bar z_{I}\}]
=\delta G^{-1}_{1}[\{z_{I},\bar z_{I}\}]
+\delta G^{-1}_{2}[\{z_{I},\bar z_{I}\}]\,.
\end{equation}
Here $\delta G^{-1}_{1}$ and $\delta G^{-1}_{2}$ have the same form as in
(\ref{delta1}), (\ref{delta2}) with $z_{I}=(T^{x}+iT^{y})/\sqrt{2S}$
parameterizing the perturbations with respect to the prescribed 
general directions $\vec\Omega_{I}$ in ${\cal H}_{MF}$.

We now consider the contributions from ${\cal H}_{fluc}$ to the energy of
the carrier ground state. In linear order in $z_{I}$ one has the 
contribution from the expectation value of $\delta G^{-1}_{1}$ 
which can be written as
\begin{equation}
E^{(1)}=\frac{1}{2}\sum_{I}
\left[\bar g_{I}z_{I}+g_{I}\bar z_{I}\right]
\label{erggradient}
\end{equation}
with $g_{I}=g^{1}_{I}+ig^{2}_{I}$ given by (\ref{g1}), (\ref{g2}).
This expression coincides with ({\ref{gradientfluc}) \cite{note1}. 
The expectation values of the carrier spin density 
$\langle\vec s(\vec r)\rangle$ entering (\ref{erggradient}) via 
the complex coefficients 
$g_{I}$ have to be computed within the ground state of the
unperturbed Hamiltonian ${\cal H}_{MF}$. This operator depends
on the Mn spin directions $\vec\Omega_{I}$ and on the
kinetic term ${\cal H}_{kin}$ for the carriers, which has not been
specified so far within this perturbational approach to the ground state
energy. Therefore the expression (\ref{erggradient})
holds formally for {\em any band representation of the carrier system}.
In particular (\ref{erggradient}) is valid for the simple parabolic-band model
as well as for more sophisticated $\vec k\cdot\vec p$ Hamiltonians.
We will come back to this important fact in section~\ref{gradient}.

There are two contribution bilinear in $z_{I}$. The first one is just the
expectation value of $\delta G^{-1}_{2}$ within the carrier ground state,
\begin{equation}
E^{(2)}_{L}=\sum_{I,J}
\bar z_{I} L_{IJ}z_{J}
\end{equation}
with
\begin{equation}
L_{IJ}=-\delta_{IJ}
 \int d^{3}rJ(\vec r-\vec R_{I})
\langle\vec s(\vec r)\rangle\vec\Omega_{I}
\end{equation}
which obviously coincides with (\ref{L}) for $\Omega_{n}=0$ and
$\vec\Omega_{I}=\vec e_{z}$ for all $I$.
The other bilinear contribution is the second-order term arising from
$\delta G^{-1}_{1}$ which reads
\begin{eqnarray}
E^{(2)}_{K} & = & \frac{S}{4}\sum_{\alpha,\beta}\Biggl[
\frac{n_{F}(\eta_{\alpha})-n_{F}(\eta_{\beta})}
{\eta_{\alpha}-\eta_{\beta}}\qquad\qquad\nonumber\\
 & & \cdot\Biggl|\sum_{I}\Big[\Re\{z_{I}\}\sin\vartheta_{I}
\left(F^{\alpha\downarrow,\beta\downarrow}
-F^{\alpha\uparrow,\beta\uparrow}\right)\nonumber\\
& & +\left(\cos\vartheta_{I}\Re\{z_{I}\}-i\Im\{z_{I}\}\right)
F^{\alpha\uparrow,\beta\downarrow}e^{-i\varphi_{I}}\nonumber\\
& & +\left(\cos\vartheta_{I}\Re\{z_{I}\}+i\Im\{z_{I}\}\right)
F^{\alpha\downarrow,\beta\uparrow}e^{+i\varphi_{I}}\Big]\Biggr|^{2}\Biggr]\,.
\label{ergK}
\end{eqnarray}
It is straightforward to see that for in the collinear case,
$\vec\Omega_{I}=\vec e_{z}$ for all $I$, this expression takes the form
\begin{equation}
E^{(2)}_{K}=\sum_{I,J}
\bar z_{I} K_{IJ}z_{J}
\end{equation}
with $K_{IJ}=K_{IJ}(\Omega_{n}=0)$ given by Eq.~(\ref{K}) \cite{note2}.

We note that the perturbational approach described here is crucially different
from the RKKY approximation often used in the theory of spin glasses
\cite{Abrikosov80}.  There the coupling of the carrier spin density to the
local moments is treated as a perturbation to the non-polarized carrier
Fermi sea. This is justified provided that the free-carrier Fermi energy
is large compared to the energy scale of the exchange coupling which
is conveniently measured in terms of the carrier mean-field splitting
$\Delta=J_{pd}SN_{Mn}$, where $N_{Mn}$ is the density of Mn spins.
However, for typical parameters of (III,Mn)V systems the Fermi energy
and $\Delta$ are fairly of the same order of magnitude with the former
quantity being often even smaller. Therefore the free-carrier ground
state is not a good starting point for perturbation theory.
To explore the true ground state of the system described by the Hamiltonian
(\ref{defmod1}) for parameters realistic for (III,Mn)V semiconductors
one should rather use the carrier ground state in the presence of a fully
polarized Mn spin system as the starting point, as done here.


\section{Instability of the collinear ferromagnetic state in the 
isotropic parabolic-band model}
\label{inst}

\subsection{General discussion of the energy gradient}
\label{gradient}

As already stressed in section \ref{pert}, the expression (\ref{gradientfluc})
obtained for the ground state energy in lowest oder in the Holstein-Primakoff
variables is very general, i.e. it is formally the same for any
representation of the carrier system (e.g. parabolic band, six-band
$\vec k\cdot\vec p$ Hamiltonian,...).

Let us now examine further the coefficients given in Eqs.~(\ref{g1}), 
(\ref{g2}) that enter the energy gradient (\ref{gradientfluc}).
The imaginary part $g_{I}^{2}$ is nonzero if and only if the vector
\begin{equation}
\vec m_{I}:=\int d^{3}rJ(\vec r-\vec R_{I})\langle\vec s(\vec r)\rangle
\end{equation}
does {\em not} lie in the plane spanned by
$\vec e_{\varphi_{I}}=(\cos\varphi_{I},\sin\varphi_{I},0)$ and
$\vec e_{z}=(0,0,1)$. Similarly, the real part $g_{I}^{1}$ is nonzero
if and only if the projection of $\vec m_{I}$ onto the plane spanned
by $\vec e_{\varphi_{I}},\vec e_{z}$ is {\em not} collinear with
the direction $\vec \Omega_{I}=(\sin\vartheta_{I}\cos\varphi_{I},
\sin\vartheta_{I}\sin\varphi_{I},\cos\vartheta_{I})$ of the Mn spin at 
site $I$.
In summary, for a given orientation of the Mn local moments to be
stationary in energy the local directions $\vec \Omega_{I}$ must
be parallel or antiparallel
with $\vec m_{I}$ at each Mn site $I$. 
An example for such a situation is the collinear ferromagnetic state 
with all Mn spins in parallel in a simple parabolic-band model. 
Here the spins of all carrier eigenstates are polarized along the
common axis of the Mn spins, and therefore $\vec m_{I}$ is always 
collinear with this direction at each site $I$.
Thus, this state is always
{\em stationary}, but, as we shall see below, in general not {\em stable}.
Moreover, the collinear ferromagnetic state turns out to be not even
stationary if anisotropy induced by spin-orbit interactions is present,
as it is the case for valence band holes in III-V semiconductors.

\subsection{Numerical results}
\label{numerical}

We now study the stability of the collinear ferromagnetic state
in a simple parabolic-band model where the carriers are characterized
by just one effective mass $m^{\ast}$. As seen before, the energy gradient
vanishes for such a state, and
the quadratic zero-frequency ($n=0$) contribution to the effective action
(\ref{fluctuation}) describes the energy of static fluctuations around
the collinear state. Thus, for this state to be stable, the matrix
$D^{-1}_{IJ}(0)$ must have non-negative eigenvalues only, while the
occurrence of negative eigenvalues of this matrix indicates that the
perfectly collinear state is not the ground state. We note that for any
arrangement of the Mn positions $R_{I}$ the matrix $D^{-1}_{IJ}(0)$ contains
a zero eigenvalue corresponding to a uniform rotation of all spins.
If $D^{-1}_{IJ}(0)$ is non-negative its eigenvalue distribution can be
interpreted as a density of states (DOS) for magnetic excitations above the
collinear state.

We have evaluated the spectrum of $D^{-1}_{IJ}(0)$ in systems given by
a simulation cube with periodic boundary conditions averaging over different
realizations of the Mn positions. The single-particle wave functions
$\psi_{\alpha\sigma}(\vec r)$ are computed in a plane-wave basis taking into
account wave vectors $\vec q$ with length up to an appropriate cutoff $q_{c}$.
The same truncated plane-wave basis is used to compute the quantities
(\ref{F}) entering (\ref{K}). Note that, for fluctuations around the
collinear ferromagnetic state, $D^{-1}_{IJ}(i\omega)$ is always real
and symmetric for real $\omega$ since all carrier wave functions 
have for a given spin projection $\sigma$ a coordinate-independent
phase (and can therefore also chosen to be real). This follows from the fact
that the single-particle Hamiltonian describes for each spin projection 
just the problem of a spinless particle in a potential landscape provided by 
the Mn ions. Since $D^{-1}_{IJ}(i\omega)$ is real and symmetric the
components of each of its eigenvectors have all the same phase
(and can be chosen to be real). Physically this corresponds to the
invariance of the system under rotations around the magnetization axis of
the collinear state.

The two upper panels of
Fig.~\ref{fig1} show results for typical system parameters for two
different values of $q_{c}$. The comparison of both panels shows that the
effects of the wave vector cutoff on the low-lying excitations have already
saturated for the smaller $q_{c}$. The by far largest contributions to the
DOS lie at positive energies, while a small amount of sightly negative
eigenvalues of $D^{-1}_{IJ}(0)$ indicate an instability of the perfectly
collinear state.

In the calculations discussed so far the Mn positions were chosen 
completely at random with uniform distribution, while in a real
(III,Mn)V semiconductor the Mn ions are supposed to be located on
the cation sites forming an fcc lattice. In the bottom panel of 
Fig.~\ref{fig1} we show data for the same system parameters as in the
top panel but with the Mn positions chosen from an appropriate fcc
lattice such that about 5 percent of all sites are occupied. Both plots
are practically identical indicating that our observations do not depend
on this detail of the modeling.

The shape of the eigenvalue distribution of the fluctuation matrix
$D^{-1}_{IJ}(0)$ is quite
sensitive in detail to the Mn density $N_{Mn}$, the carrier density $p$,
and the Hamiltonian parameters $m^{\ast}$, $J_{pd}$, $a_{0}$. In our numerics,
we have extensively investigated this high-dimensional parameter space in
regions realistic for (III,Mn)V semiconductors. The general finding is
that {\em the ground state of the system is generically noncollinear}.
To our observation there occur always negative eigenvalues
of $D^{-1}_{IJ}(0)$ indicating the instability of the collinear state,
provided that large enough wave vector cutoffs and system sizes are
considered.

To analyze further the nature of this instability we consider the
participation ratio
\begin{equation}
p(E)=\left[NV\sum_{I}|\alpha_{I}(E)|^{4}\right]^{-1}
\label{partrat}
\end{equation}
where $\alpha_{I}(E)$ is the $I$-th component of the (normalized)
eigenvector of $D^{-1}_{IJ}(0)$ with eigenvalue $E$, and the summation goes
over all $NV$ Mn sites. This quantity is an estimate for the fraction of
components of $\alpha(E)$ being substantially nonzero. For instance, if
a vector contains exactly a fraction of $p$ nonzero components of equal
modulus and all others being zero, its participation ratio is $p$.
The largest participation ratio of unity is achieved for the zero-energy
uniform rotation mode where all components of the corresponding
eigenvector are equal.

Fig.~\ref{fig2} shows the disorder-averaged participation ratio for the same 
situation as in the top panel of Fig.~\ref{fig1}.
The negative-energy modes have clearly higher participation ratio than
the eigenvectors at positive energy. This shows that {\em the instability of 
collinear state is due to long-ranged dynamics involving a large fraction
of the spins present in the system}. Qualitatively the same observations
are made for other values of system parameters.


\section{Noncollinear ground states and the influence of a magnetic field}
\label{noncollmag}

In the previous section we have investigated the stability of the
collinear ferromagnetic state in the parabolic-band kinetic-exchange model.
This state is always stationary (i.e. has a vanishing energy gradient)
but not necessarily stable. In the present section we extend 
our ground state studies using the energy gradient expression
(\ref{gradientfluc}). We will consider the case of simple
parabolic bands as well as more sophisticated kinetic carrier
Hamiltonians incorporating spin-orbit anisotropy.

\subsection{Stationary states}

\subsubsection{Helical states in the parabolic-band model}
\label{helical}

We now discuss a particular class of metastable states in the 
parabolic-band model.

The energy gradient expression (\ref{gradientfluc}) can be employed in a 
numerical steepest-descent procedure to search for true energy minima.
To this end, one starts with the collinear state with all spins
pointing into the $z$-direction
and steps down in energy
by performing sufficiently small rotations of the Mn spins according
to some negative eigenvector of $D^{-1}_{IJ}(0)$ (or some linear
combination of them). Now consider the case that 
only one eigenvector is involved with all components having a
certain common phase. The resulting orientations of Mn spins will all lie
in the same plane spanned by the $z$-axis and a direction in the
$xy$-plane that is determined by the above phase factor.
Since the parabolic-band Hamiltonian is invariant under spin-flips
with respect to this given plane, the local spin density 
$\langle\vec s(\vec r)\rangle$ and consequently the vectors
$\vec m_{I}$ will also lie in this plane \cite{note3}.
As seen in the previous subsection, for such a situation
the imaginary part $g_{I}^{2}$ of the coefficients 
vanishes identically for all $I$.
When now applying the energy gradient expression in a steepest-decent
procedure the real part $g_{I}^{1}$ of the gradient coefficients 
will just move the directions $\vec \Omega_{I}$ within this plane, 
while the imaginary
parts $g_{I}^{2}$ remain strictly zero. Eventually this procedure will end up
in a stationary  state with all Mn spins lying in a plane 
prescribed by the initial departure from the collinear state.

These conclusions are confirmed by explicit numerics where
such {\em complanar} (or {\em helical}) stationary states are indeed observed.
This states can be seen as {\em bona fide} local minima
on the energy landscape.

However, as explained in more detail in the following subsection,
such type of energy minima occur only in isotropic models but not
in systems with spin-orbit anisotropy.

\subsubsection{Nonstationarity of the collinear state in the presence of 
spin-orbit anisotropy}
\label{nonstatsixband}

We now examine the energy gradient (\ref{gradientfluc}) with respect to
a six-band $\vec k\cdot\vec p$ Hamiltonian 
involving spin-orbit anisotropy \cite{Luttinger55,Abolfath01,Chow99}.
In this case the rotational invariance in spin space 
is broken down to the cubic symmetry of the underlying GaAs crystal.

Let us first consider the collinear ferromagnetic state.
In this case we find numerically that the vectors $\vec m_{I}$ are 
{\em not parallel} (or antiparallel)
with any given common orientation of the Mn spins.
This finding includes also the crystallographic symmetry axes
(1,0,0), (1,1,0), and (1,1,1) or their equivalents.

There are two ways to restore the collinearity between the vectors
$\vec m_{I}$ and a common orientation of the Mn spins:

(i) {\em Virtual crystal or continuum approximation:} If the Mn spins
are approximated as an continuum and  
point all along one of the crystallographic symmetry axes
(1,0,0), (1,1,0), and (1,1,1) or their equivalents, 
the $\vec m_{I}$ are always antiparallel to this direction.
This finding explicitly confirms an assumption which earlier
spin-wave calculations in the six-band continuum model were based on
\cite{Konig01b}. Formally the continuum limit is reached by putting the
parameter $a_{0}$ in the coupling function (\ref{excpl}) to large
values. In this limit the disorder with respect to the Mn positions
is removed.

(ii) {\em Zero spin-orbit coupling:} For vanishing spin-orbit coupling
the vectors $\vec m_{I}$ are antiparallel to any given common
direction of the Mn spins.

Thus, in the presence of disorder with respect to the Mn positions
and a finite anisotropy induced by spin-orbit interaction, 
{\em the collinear ferromagnetic
is never stationary and therefore not the ground state}. 
This statement includes the case that the common direction of the Mn spins
is along some crystallographic symmetry axis.

This important result resembles recent findings by Zarand and Janko
\cite{Zarand01} obtained within the RKKY approximation, where the
ground state of the system was also found to be noncollinear
(or, as termed there,  frustrated).

Finally we briefly remark on the case that the directions of 
all Mn spins lie in some common plane. According to our numerics, and 
as to be expected from the above results, the
local spin density $\langle\vec s(\vec r)\rangle$ and consequently the vectors
$\vec m_{I}$ do not lie in plane of the Mn spins. 
Therefore truly {\em complanar} states being local energy minima 
do not exist if spin-orbit anisotropy is present

\subsection{Steepest-descent results and the role of an external field
in the parabolic-band model}
\label{stepdesc}

We have employed the energy gradient expression (\ref{gradientfluc}) in a 
numerical steepest-descent procedure outlined in subsection \ref{helical}
to search for stationary states in the parabolic-band model.
These states can be considered as {\em bona fide} energy minima. 
Our results are as follows: In cases where the energy minimum found by this
method is close to the collinear state (with a magnetization of about
90 percent of the maximum value or more), this minimum appears to be
unique (for a given disorder realization) and can therefore be considered as 
the true absolute ground state of the system. However, in situations 
where the magnetization is reduced more substantially (by, say, about 20
percent or more) the energy minima found are not unique anymore but
depend on technical details of the numerical procedure. In such cases the
system is essentially spin-glass like with a complicated energy
landscape. This situation occurs typically at larger density ratios
$p/N_{Mn}$ of carriers and Mn spins. For the system shown in 
Fig.~\ref{fig1} for
instance the magnetization values in the energy minima typically found are
of about 30 to 40 percent of the collinear state.

Finally we consider the role of a magnetic field coupled to the Mn spins,
\begin{equation}
{\cal H}_{B}=\vec\Delta\cdot\sum_{I}\vec\Omega_{I}
\end{equation}
where $\vec\Delta=g\mu_{B}S\vec B$ is the vector of Zeeman couplings.
Here the magnetic field is coupled 
only to the Mn spins and not to the band carriers, since their mean-field
spin splitting $\Delta=J_{pd}SN_{Mn}$ is large compared to the Zeeman
couplings considered here. Therefore this contribution to the effective
carrier spin splitting is negligible, and one should expect the collinear
state to be stabilized by an external Zeeman field which is equal in
magnitude to the smallest (negative) eigenvalue of the fluctuation matrix
$D_{IJ}^{-1}(0)$. For instance, from the data shown in Fig~\ref{fig1}
this Zeeman field would be about $0.5{\rm meV}$.

A magnetic field in the $z$-direction leads to an additional
contribution to the gradient components $g_{I}^{1}$ of
$\Delta^{z}\sqrt{2/S}\sin\vartheta_{I}$, which we shall consider in the 
following. We employ the following numerical procedure.
Starting from the collinear state with all Mn spins pointing in the 
$z$-direction we obtain a noncollinear energy minimum by the 
steepest-descent method described above.
Then we add a magnetic field in the $z$-direction and an repeat the
procedure to reach a new energy minimum. This state is then the new starting 
point for a steepest-descent walk with an increased magnetic field.
By iterating this procedure we obtain a zero-temperature magnetization curve
as a function of the external field.

Our numerical results are as follows: In cases where the zero-field
ground state magnetization is substantially reduced from its maximum value
for the collinear state (i.e. in the ``glassy'' regime), the magnetization
at finite field depends quantitatively on the step width that is used 
in increasing the field. Fig.~\ref{fig3} shows magnetization data for the
same system parameters as in the data before. As seen from there the
Zeeman field needed to realign all spin along the $z$-direction is 
somewhat larger than the expected value of $\Delta^{z}=0.5{\rm meV}$,
which is a typical hysteresis effect.


\section{The small-frequency dependence of the action kernel
and adiabatic dynamics}
\label{finfreq}

In section \ref{inst} we studied the stability of the
collinear ferromagnetic state within a parabolic band model in terms
of the static component of the effective action kernel
$D^{-1}_{IJ}$. Here we extend these analysis to the regime of small
but finite Matsubara frequencies $\Omega_{n}$.

Expanding $K_{IJ}(\Omega_{n})$ around $\Omega_{n}=0$ one has
\begin{equation}
K_{IJ}(\Omega_{n})=K_{IJ}(0)+i\Omega_{n}K'_{IJ}+\cdots
\end{equation}
with
\begin{equation}
K'_{IJ}=-\frac{S}{2}\sum_{\alpha,\beta}\left[
\frac{n_{F}(\eta_{\alpha})-n_{F}(\eta_{\beta})}
{(\eta_{\alpha}-\eta_{\beta})^{2}}
F^{\alpha\downarrow,\beta\uparrow}_{I}
F^{\beta\uparrow,\alpha\downarrow}_{J}\right]\,.
\label{K'}
\end{equation}

The matrix $K'$ fulfills an important sum rule which we derive now.
As mentioned before, for the collinear state with all Mn spins pointing
in the $z$ direction all carrier eigenstates have either spin up or spin down,
and one ends up with two separated problems for spinless fermions in a 
potential landscape. The stationary Schr\"odinger equations for spin down 
and spin up particles read
\begin{eqnarray}
-\frac{\hbar^{2}\nabla^{2}}{2m^{\ast}}\psi_{\alpha\downarrow}(\vec r)
-\frac{S}{2}\sum_{I}\int d^{3}rJ(\vec r-\vec R_{I})
\psi_{\alpha\downarrow}(\vec r) ) & & \nonumber\\
=\varepsilon_{\alpha}\psi_{\alpha\downarrow}(\vec r)\,, & & \\
-\frac{\hbar^{2}\nabla^{2}}{2m^{\ast}}\psi_{\beta\uparrow}(\vec r)
+\frac{S}{2}\sum_{I}\int d^{3}rJ(\vec r-\vec R_{I})
\psi_{\beta\uparrow}(\vec r) & & \nonumber\\
=\varepsilon_{\beta}\psi_{\beta\uparrow}(\vec r)\,. & & 
\end{eqnarray}
Taking all wave functions to be real and combining these equations one finds
\begin{equation}
\left(\eta_{\beta}-\eta_{\alpha}\right)
\langle\psi_{\alpha\downarrow}|\psi_{\beta\uparrow}\rangle
=S\sum_{I}F^{\alpha\downarrow,\beta\uparrow}_{I}
\end{equation}
where $\langle\cdot|\cdot\rangle$ denotes a scalar product 
between spinless wave functions. Since both the spin up and the spin down
carrier wave functions fully span the Hilbert space 
of a single spinless particle we have
\begin{eqnarray}
\sum_{\beta}\langle\psi_{\alpha\downarrow}|\psi_{\beta\uparrow}\rangle
\langle\psi_{\beta\uparrow}|\psi_{\alpha\downarrow}\rangle
=\langle\psi_{\alpha\downarrow}|\psi_{\alpha\downarrow}\rangle=1
& & \nonumber\\
=\sum_{\beta}\frac{S^{2}}{\left(\eta_{\beta}-\eta_{\alpha}\right)^{2}}
\sum_{IJ}F^{\alpha\downarrow,\beta\uparrow}_{I}
F^{\beta\uparrow,\alpha\downarrow}_{J}
\label{sum1}
\end{eqnarray}
and similarly
\begin{equation}
\sum_{\alpha}\frac{S^{2}}{\left(\eta_{\beta}-\eta_{\alpha}\right)^{2}}
\sum_{IJ}F^{\alpha\downarrow,\beta\uparrow}_{I}
F^{\beta\uparrow,\alpha\downarrow}_{J}=1\,.
\label{sum2}
\end{equation}
From these sum rules one derives \cite{note4}
\begin{equation}
\sum_{IJ}K'_{IJ}=\frac{1}{2S}\left(n_{\uparrow}-n_{\downarrow}\right)
\label{sumK'}
\end{equation}
where $n_{\uparrow}$, $n_{\downarrow}$ are the numbers of up and down spin
carriers, respectively. 

We now consider the limit of large inverse temperature $\beta$ where the
bosonic Matsubara frequencies $\Omega_{n}=2n\pi/\beta$ can be treated as a
continuous variable $\Omega$. We are interested in the adiabatic limit of
the spin dynamics which is described by the low-frequency 
limit of the effective action (\ref{fluctuation}).
In the expansion of the Fourier transformed Holstein-Primakoff variables,
\begin{equation}
z_{I}(\Omega)=z_{I}(0)+\Omega
\left(\frac{d z_{I}}{d\Omega}\right)_{\Omega=0}+\cdots
\end{equation}
the zero-frequency component $z_{I}(0)$ corresponds to Holstein-Primakoff
variables constant in imaginary time $\tau$. Here we consider the uniform 
case $z_{I}(\tau)=\sqrt{S\eta}\exp(i\chi)$ for all $I$. This describes
a uniform rotation of all Mn spins around the axis 
$(\sin\chi,-\cos\chi,0)$ by an angle parametrized by $\eta$. 
As seen in section \ref{inst} the matrix $D^{-1}_{IJ}(0)$ annihilates
the vector of the zero-frequency components 
$z_{I}(\Omega=0)=\beta\sqrt{S\eta}\exp(i\chi)$. Therefore, using the sum rule 
(\ref{sumK'}), the low-frequency expansion of the effective action 
(\ref{fluctuation}) reads
\begin{equation}
S_{\rm fluc}^{(2)}=\beta(i\Omega)\eta
\left(-SN_{Mn}V+\frac{1}{2}\left(n_{\uparrow}-n_{\downarrow}\right)\right)
+\cdots
\end{equation}
where $N_{Mn}V$ is the number of Mn ions in the system.
With an analytical continuation to real time, the above expression linear
in $\Omega$ is the geometric phase generated by an adiabatic uniform
rotation of all Mn spins (pointing initially along the $z$-direction)
around an axis in the $xy$-plane 
by an angle $\beta\Omega\eta$. The second term 
in the parentheses stems from the carriers which adiabatically 
follow the Mn spins. Note that this 
contribution comes with a different sign.
This is due to the fact that we are dealing with an effective action for 
the Mn spins only where the carriers have been integrated out,
and therefore only the Mn spins are 'actively' rotated.
In our formalism the initial polarization axis of the Mn spins defines the
quantization axis for the carrier spins, and rotating this axis
is just a 'passive' rotation of the carrier spin coordinate system.
Therefore the geometric phase stemming from the carriers has a different sign.

We now turn to the case of general (i.e. non-uniform) 
adiabatic rotations of the Mn spins parametrized by other 
eigenvectors of $D^{-1}_{IJ}(0)$ than the one discussed above.
Then the geometric phase of the localized spins still 
stems from the frequency-dependent part of the 
diagonal matrix $L_{IJ}(\Omega)$ (cf.~Eq.~(\ref{L})) while 
the largest contribution to the carrier phase will arise from 
$D^{-1}_{IJ}(0)$ itself and 
the derivative $K'_{IJ}$. We therefore expect $K'_{IJ}$ to be dominantly 
diagonal when expressed in the eigenbasis of $D^{-1}_{IJ}(0)$.
In fact, this expectation is confirmed by numerical evaluations of 
this quantity. As a measure for the 'diagonality' of a given matrix $A$ we
consider
\begin{equation}
D(A)=\left(\frac{\sum_{i}A^{2}_{ii}}{{\rm tr}A^{2}}\right)^{1/2}\,.
\end{equation}
This quantity is unity if $A$ is diagonal, and of order the inverse of the
square root of the dimension of $A$ if all of its elements are of the same 
order of magnitude.

We have evaluated the 'diagonality' $D$ of $K'_{IJ}$
expressed in the eigenbasis of $D^{-1}_{IJ}(0)$ for various sets of system
parameters and averaged this quantity over many disorder realizations
with respect to the Mn positions in space. We consistently find
$D$ substantially larger than the average value of a general matrix.
For instance, for the parameters used in Figs.~\ref{fig1},\ref{fig2}
we find $\bar D=0.560$ (average over all disorder realizations) with
a fluctuation $\Delta D=\sqrt{{\overline{D^{2}}}-\bar D^{2}}=0.045$.
This is by an order of magnitude larger than the value of a general
matrix of this dimension which is $D=1/\sqrt{400}=0.05$.


\section{Conclusions}
\label{conclusions}

We have studied the ground state properties of kinetic-exchange models for
carrier-induced ferromagnetism in 
(III,Mn)V semiconductors with randomly distributed Mn ions. Our method
is embedded in a path integral spin-wave type
formalism leading to an effective action for the Mn spins with full 
Matsubara frequency dependence. The zero-frequency contribution to
this action is equivalent to static perturbation theory and characterizes
the stability of a given spin configuration, while the component linear in 
frequency can be interpreted as the joint Berry phase of the Mn and carrier
system.

Our perturbational approach to the ground state of the system studied here
differs from the RKKY approximation in so far as we not do do perturbation
theory around the free-carrier ground state but around the carrier ground 
state in the presence of a fully aligned collinear Mn spin system.
This is appropriate since the mean-field spin splitting is for
realistic system parameters {\em not} small compared to the Fermi energy.
Therefore it cannot be regarded as a small perturbation to the free-carrier
ground state, as done in the RKKY approximation. This approximation
actually works well for metallic spin glass systems where the coupling to
local moments is indeed a small perturbation \cite{Abrikosov80}.

For parabolic-band carriers the collinear ferromagnetic state 
with all Mn spins in parallel is always {\em stationary} but generically
{\em unstable}. This instability can be characterized in terms of inverse
participation ratios and is due to long-ranged nonlocal spin fluctuations.
We also have presented results for the ground state magnetization as a function
of an external field.

For carrier dispersions involving anisotropy induced by spin-orbit coupling
the collinear state is not even stationary and therefore also not the
ground state. This interplay between the anisotropy in the
carrier system and the disorder in the Mn positions reflects recent findings
by Zarand and Janko \cite{Zarand01} obtained within the RKKY approximation.
The stationarity of the collinear state is restored in the continuum or
{\em virtual crystal} approximation where disorder is neglected. 

\acknowledgements{I thank A.~L. Chudnovskiy, P.~H. Dederichs, J. K\"onig, 
Q. Niu, F. von Oppen, D. Pfannkuche, N. Samarth, P. Schiffer,
C. Timm, and, in particular, A.~H. MacDonald for useful discussions.
This work has been supported by the Swiss NSF, 
DARPA, ARO, the Indiana 21st Century fund, and the Welch Foundation.}


%
\begin{figure}
\centerline{\includegraphics[width=8cm]{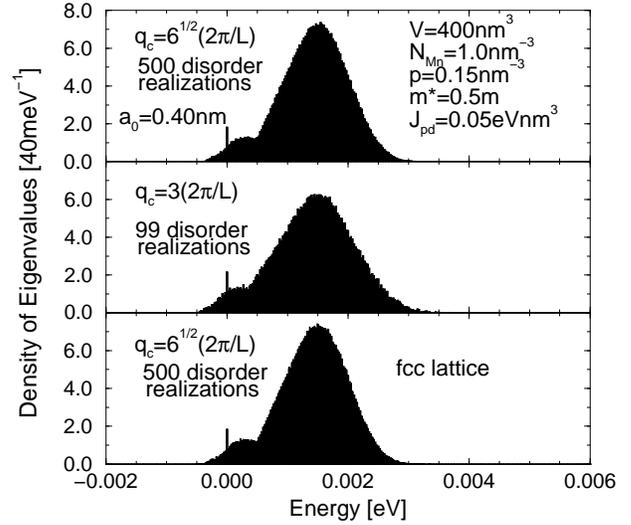}}
\caption{The disorder-averaged density of states of magnetic excitations 
for a simulation cube of volume $V=L^{3}=400{\rm nm}^{3}$ with a Mn density 
of $N_{Mn}=1.0{\rm nm}^{-3}$ 
and a density of $p=0.15{\rm nm}^{-3}$ of carriers having a band mass
of half the bare electron mass. The strength of the exchange interaction
between ions and carriers is $J_{pd}=0.05{\rm eVnm}^{-3}$ with a spatial
range of $a_{0}=0.40{\rm nm}$ The two upper panels show data for different
wave vector cutoff $q_{c}$ with the Mn positions chosen completely at 
random. The lowest panel contains data for the same situation as the
top one but with the Mn positions chosen from an fcc lattice.
The peaks at zero energy are due to the uniform 
rotation mode which strictly occurs in any disorder realization.
\label{fig1}}
\end{figure}
\begin{figure}
\centerline{\includegraphics[width=8cm]{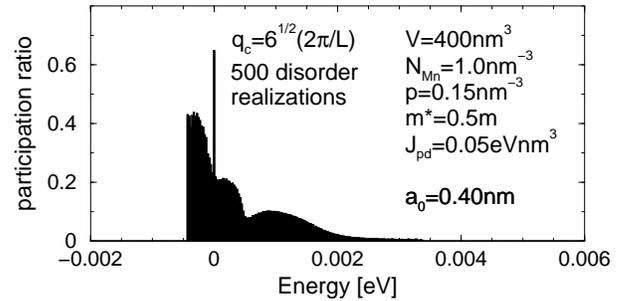}}
\caption{The disorder-averaged participation ratio for the same situation as
in the top panel of Fig.~\protect{\ref{fig1}}. The data is averaged over the
sample intervals of the histogram. The value at zero energy is enhanced 
due to the contribution of the uniform rotation mode in that sample
interval. 
\label{fig2}}
\end{figure}
\begin{figure}
\centerline{\includegraphics[width=8cm]{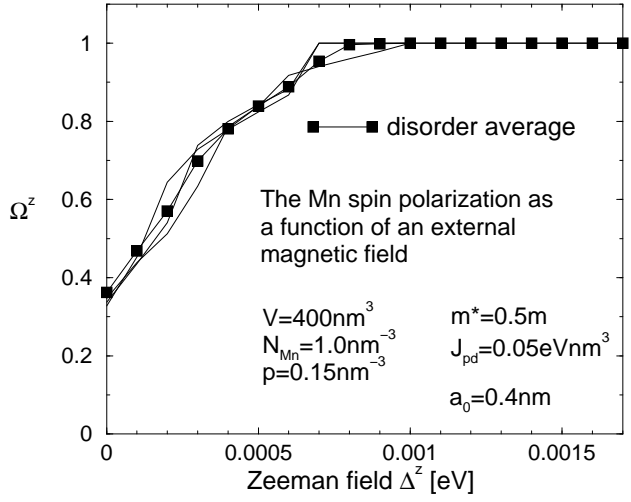}}
\caption{The $z$-component of the Mn spin polarization
$\Omega^{z}=\sum_{I}\Omega^{z}_{I}/(N_{Mn}V)$ as a function of an external
magnetic field for the same parameters as in the top 
panel of Fig.~\protect{\ref{fig1}}. The Zeeman coupling is increased in units
$0.1{\rm meV}$. The thin lines are results for three individual disorder
realizations while the thick graph is the disorder 
average over 11 realizations. 
\label{fig3}}
\end{figure}

\begin{references}
\bibitem{note1}
Note that the expression Eq.~(\ref{erggradient}) has, 
just as $g_{I}$, dimension
of energy, since the $z_{I}$ there are dimensionless. This is different from
(\ref{gradientfluc}) where the quantities $z_{I}(\Omega_{n})$ have the
dimension of the inverse temperature such that ${\cal S}_{\rm fluc}^{(1)}$
is dimensionless.
\bibitem{note2}
We note that the summation over $\alpha$ and $\beta$ in Eq.~(\ref{ergK})
is unrestricted, i.e. is performed 
over all single-particle states. If one would
restrict the summation to,  say, states $\alpha$ that are occupied in the
ground state and states $\beta$ that are unoccupied, a factor of 
$2$ would occur.
\bibitem{note3}
If the spin flip is performed with respect to the $xz$-plane
the corresponding operator acting on the carrier states is just the 
operator $\cal K$ of complex conjugation which changes the sign of
$\sigma^{y}$ but leaves the other Pauli matrices invariant. In the case
of the $xy$-plane or $yz$-plane the spin flip operator 
is $\sigma^{x}\cal K$ and $\sigma^{z}\cal K$, respectively.
For a general orientation of this plane one has to use the 
appropriate linear combination of these operators.
If all Mn spins lie in such a plane the parabolic-band Hamiltonian
is invariant under this spin flip operation.
\bibitem{note4}
Note that the sum rules Eqs.~(\ref{sum1}), (\ref{sum2}) are independent  
of our choice to take all wave functions to be real since
possible common phase in the up or down spins cancel in each term
in the summation over $I,J$.
\end{references}
\end{document}